\documentclass[conference]{IEEEtran}
\usepackage{cite}
\usepackage{amsmath,amssymb,amsfonts}
\usepackage{algorithmic}
\usepackage{graphicx}
\usepackage{textcomp}
\usepackage{xcolor}
\usepackage{booktabs}
\usepackage{multirow}
\usepackage{array}
\usepackage{algorithm}
\usepackage{tabularx}
\usepackage{tikz}
\usepackage{pgfplots}
\pgfplotsset{compat=1.17}
\usetikzlibrary{shapes,arrows,positioning,calc,decorations.pathreplacing}

\def\BibTeX{{\rm B\kern-.05em{\sc i\kern-.025em b}\kern-.08em T\kern-.1667em\lower.7ex\hbox{E}\kern-.125emX}}

\begin{document}

\title{A Neuro-Symbolic System for Interpretable Multimodal Physiological Signals Integration in Human Fatigue Detection}

\author{\IEEEauthorblockN{Mohammadreza Jamalifard,
Yaxiong Lei,
Parasto Azizinezhad,\\
Javier Fumanal Idocin, and 
Javier Andreu-Perez}
\IEEEauthorblockA{School of Computer Science and Electronic Engineering\\
University of Essex, United Kingdom\\
Email: \{m.jamalifard, yaxiong.lei, p.azizinezhad, j.fumanal-idocin, j.andreu-perez\}@essex.ac.uk}
}
\maketitle

\begin{abstract}
We propose a neuro-symbolic architecture that learns four interpretable physiological concepts, \textit{oculomotor dynamics, gaze stability, prefrontal hemodynamics, and multimodal}, from eye-tracking and neural hemodynamics, functional near-infrared spectroscopy, (fNIRS) windows using attention-based encoders, and combines them with differentiable approximate reasoning rules using learned weights and soft thresholds, to address both rigid hand-crafted rules and the lack of subject-level alignment diagnostics. We apply this system to fatigue classification from multimodal physiological signals, a domain that requires models that are accurate and interpretable, with internal reasoning that can be inspected for safety-critical use. In leave-one-subject-out evaluation on 18 participants (560 samples), the method achieves 72.1\% $\pm$ 12.3\% accuracy, comparable to tuned baselines while exposing concept activations and rule firing strengths. Ablations indicate gains from participant-specific calibration (+5.2 pp), a modest drop without the fNIRS concept (-1.2 pp), and slightly better performance with {\L}ukasiewicz operators than product (+0.9 pp). We also introduce \emph{concept fidelity}, an offline per-subject audit metric from held-out labels, which correlates strongly with per-subject accuracy ($r=0.843$, $p<10^{-4}$).
\end{abstract}



\section{Introduction}

We introduce a neurosymbolic system designed to address the lack of mechanisms supporting \emph{subject-level auditing and troubleshooting} in models applied to noisy or distribution-shifted multimodal physiological signals~\cite{kakhi2025fatigue}. Despite advances in multimodal sensing (e.g., eye tracking, EEG, fNIRS) \cite{lei5732154comprehensive}, current deep learning approaches are limited by their susceptibility to inter-subject variability and struggle with real-world generalization~\cite{albadawi2022review,lei2025quantifying, lei2023end}. We apply this system to the problem of fatigue, which substantially impairs human performance in safety-critical domains including transportation, healthcare, and industrial operations~\cite{caldwell2019fatigue}.

First, interpretable fatigue models typically rely on hand-crafted rules inspired by literature~\cite{lei2023end}. However, inter-individual and context-dependent variations often alter the magnitude, and even the direction of physiological responses, rendering fixed rule sets may be unreliable across different subjects and tasks~\cite{lohani2019review}. This motivates data-adaptive concept definitions and thresholds.

We address these limitations through a neuro-symbolic framework with three key contributions:

\begin{enumerate}
\item We propose an attention-based concept extraction architecture that learns four interpretable physiological concepts including \textit{oculomotor dynamics, gaze stability, prefrontal hemodynamics, and multimodal}, from eye-tracking
and fNIRS features using data-adaptive thresholds explained in the methodology.

\item We introduce a differentiable approximate reasoning layer as a structured interpretability module, with transparent fixed rule templates and end-to-end learned rule weights and soft thresholds that adapt to the data distribution.

\item We conduct extensive ablation studies on normalization, concept importance, logic operator variants, and learned vs.\ fixed thresholds, and report concept fidelity as a reliability metric ($r=0.843$ correlation with accuracy).
\end{enumerate}

\section{Related Work}

\subsection{Physiological Fatigue Detection}

Multimodal approaches combining oculomotor and neuroimaging signals have shown promise for fatigue assessment. Eye-tracking features including oculomotor dynamics, blink patterns, and gaze stability have been associated with alertness states~\cite{ajayi2025multimodal}. Reviews highlight a range of oculomotor metrics including saccade velocity, fixation patterns, and gaze dispersion as indicators of fatigue-related performance degradation~\cite{kashevnik2024gaze}. However, the relative contribution of specific eye-tracking features varies across studies and populations, motivating data-driven feature weighting rather than a priori assumptions about indicator importance.
fNIRS captures prefrontal hemodynamic changes associated with sustained attention and fatigue-related resource changes~\cite{ayaz2012optical}. The prefrontal cortex shows increased oxygenated hemoglobin concentration under higher cognitive workload~\cite{causse2017mental}. However, cross-subject generalization remains challenging due to substantial individual differences~\cite{hossain2023bci}. Relatively fewer works jointly model eye tracking and fNIRS for fatigue assessment under cross-subject evaluation settings.

\subsection{Neuro-Symbolic AI and Approximate Reasoning}

Neuro-symbolic systems integrate neural pattern recognition with symbolic reasoning, providing both learning flexibility and interpretability~\cite{marra2024nesy}. A recent systematic review found research concentrated in learning/inference (63\%) and logic/reasoning (35\%), with gaps in explainability~\cite{colelough2025nesy}. Logic Tensor Networks enable differentiable reasoning through fuzzy semantics~\cite{badreddine2021logic}. Concept Bottleneck Models (CBMs) constrain predictions through human-interpretable intermediate representations~\cite{koh2020concept}, with recent extensions including post-hoc CBMs~\cite{yuksekgonul2022posthoc}. Fuzzy logic provides a natural framework for physiological computing where concepts exist on continuous spectra~\cite{kaya2024eeg, fumanal2024ex}. Recent surveys highlight growing interest in neuro-symbolic approaches for safety-critical applications~\cite{bhuyan2024nesy}.

\subsection{Uncertainty and Reliability in Classification}

For high-stakes decisions, interpretable models are inherently preferable to post-hoc explanations~\cite{rudin2019stop}. Model calibration is essential for trustworthy deployment~\cite{guo2017calibration}. MC Dropout~\cite{gal2016dropout} and deep ensembles~\cite{lakshminarayanan2017simple} provide prediction-level uncertainty, but do not identify systematic subject-level factors affecting reliability. Individual differences in physiological responses pose fundamental challenges for cross-subject generalization~\cite{fairclough2009physiological}. Our concept fidelity metric addresses this gap by quantifying how well learned representations capture discriminative information for each individual.

\section{Experimental Setup}

\subsection{Data Collection}

We collected multimodal physiological data from 18 healthy adults (10 female, 8 male; age $27.7 \pm 6.5$) during an ethics-approved fatigue protocol. Participants provided written consent and reported normal/corrected-to-normal vision with no epilepsy, neurological/psychiatric disorders, or skin allergies. The 40–60 minute session occurred in a controlled lab using a chin-rest for stabilization. Standard eye-tracker calibration and drift correction were performed.

The experimental session comprised three phases: (1) baseline assessment: A battery of oculomotor tasks including pro-saccade, anti-saccade, and smooth pursuit tasks; we use eye tracking here as a sensing modality for window-level oculomotor features rather than for fine-grained cognitive inference. (2) fatigue induction: A high-load sequence consisting of sustained visual search and mental arithmetic (approximately 30 minutes) designed to deplete cognitive resources; and (3) post-task assessment: A repetition of the baseline oculomotor battery to quantify performance degradation and pattern changes.

\begin{table}[t]
\centering
\caption{Acquisition and preprocessing summary}
\label{tab:data_summary}
\footnotesize
\setlength{\tabcolsep}{3pt}
\renewcommand{\arraystretch}{1.1}
\resizebox{0.9\columnwidth}{!}{%
\begin{tabularx}{\linewidth}{l X}
\toprule
\textbf{Component} & \textbf{Core details} \\
\midrule
\textbf{Eye-tracking} & EyeLink 1000 Plus (2000\,Hz); timestamps reconstructed to remove LSL jitter/duplicates. \\
\textbf{fNIRS} & 8-channel CW; bandpass 0.01--0.2\,Hz; flat channels ($\sigma<10^{-10}$) removed. \\
\textbf{Pupil preprocessing} & Blink detection at $2.5\sigma$ below median, 50\,ms dilation, linear interpolation; bandpass 0.01--4.0\,Hz. \\
\textbf{Windows} & Modalities aligned to 10\,Hz; 10\,s windows with 50\% overlap; eye features computed at native rate then aggregated. \\
\textbf{Labels} & Alert (baseline) vs.\ Fatigued (post-induction); 560 samples (280/280). \\
\bottomrule
\end{tabularx}
}
\end{table}

We employed leave-one-subject-out cross-validation (LOSO-CV), training on 17 subjects and evaluating on the held-out subject in each of 18 folds. Main results (Table~\ref{tab:results}) use three random seeds (42, 123, 456) to align computational cost across methods including nested CV tuning. Ablation studies (Tables~\ref{tab:ablation_norm}--\ref{tab:ablation_polarity}) were conducted under the calibration configuration with the same three seeds. We verified key findings are consistent with five-seed runs (72.1\% $\pm$ 11.8\% for NeSy). All baseline methods received identical preprocessed features with the same participant-aware normalization to ensure fair comparison.

\subsection{Baseline Methods}

We compare against an ablated model (no logic layer) and standard ML baselines. Baselines are evaluated with (i) scikit-learn defaults and (ii) modest nested-CV tuning (inner-fold grid search) to avoid optimistic bias under LOSO. Normalization is performed within each fold to prevent leakage.

\section{Methods}

\subsection{Problem Formulation}

\begin{figure*}[t]
\centering
\resizebox{\textwidth}{!}{%
\begin{tikzpicture}[
    scale=0.92, transform shape,
    node distance=0.32cm,
    input/.style={rectangle, draw=gray!70, fill=gray!10, text width=1.9cm, text centered, minimum height=0.72cm, font=\scriptsize, rounded corners=2pt, line width=0.7pt},
    normblock/.style={rectangle, draw=orange!70, fill=orange!10, text width=1.9cm, text centered, minimum height=0.72cm, font=\scriptsize, rounded corners=2pt, line width=0.7pt},
    encoder/.style={rectangle, draw=blue!70, fill=blue!10, text width=2.2cm, text centered, minimum height=0.78cm, font=\scriptsize, rounded corners=2pt, line width=0.7pt},
    concept/.style={rectangle, draw=green!60!black, fill=green!10, text width=2.35cm, text centered, minimum height=0.62cm, font=\scriptsize, rounded corners=2pt, line width=0.7pt},
    rule/.style={rectangle, draw=purple!70, fill=purple!10, text width=2.55cm, text centered, minimum height=0.60cm, font=\scriptsize, rounded corners=2pt, line width=0.7pt},
    aggregate/.style={rectangle, draw=red!70, fill=red!10, text width=1.6cm, text centered, minimum height=0.70cm, font=\scriptsize, rounded corners=2pt, line width=0.7pt},
    output/.style={rectangle, draw=red!80!black, fill=red!15, text width=1.6cm, text centered, minimum height=0.78cm, font=\scriptsize, rounded corners=3pt, line width=0.9pt},
    arrow/.style={->, >=stealth, line width=0.9pt, draw=gray!70},
    dataarrow/.style={->, >=stealth, line width=1.05pt, draw=blue!50},
    label/.style={font=\scriptsize\bfseries, text=gray!70},
    sublabel/.style={font=\tiny, text=gray!60},
]

\node[input] (eye) {Eye-tracking\\{\tiny 42 features}};
\node[input, below=0.65cm of eye] (fnirs) {fNIRS\\{\tiny 48 features}};

\node[sublabel, below=0.02cm of eye.south, anchor=north] {\tiny pupil, gaze, blink};
\node[sublabel, below=0.02cm of fnirs.south, anchor=north] {\tiny HbO, HbR, 8 ch};

\node[normblock, right=0.85cm of eye] (norm_eye) {Participant\\Norm};
\node[normblock, right=0.85cm of fnirs] (norm_fnirs) {Participant\\Norm};

\node[encoder, right=0.85cm of norm_eye] (eye_enc) {Eye Encoder\\{\tiny Attn + MLP}};
\node[encoder, right=0.85cm of norm_fnirs] (fnirs_enc) {fNIRS Encoder\\{\tiny Attn + MLP}};

\node[concept, right=0.95cm of eye_enc, yshift=0.18cm] (c1) {$C_1$: Oculomotor};
\node[concept, below=0.18cm of c1] (c2) {$C_2$: Gaze-Vig.};
\node[concept, right=0.95cm of fnirs_enc, yshift=0.18cm] (c3) {$C_3$: Prefrontal};
\node[concept, below=0.18cm of c3] (c4) {$C_4$: Multimodal};

\node[rule, right=1.05cm of c1, yshift=-0.45cm] (r1) {$f_1$: $\beta_1(\tilde{C}_1)$};
\node[rule, below=0.18cm of r1] (r2) {$f_2$: $\beta_2(\tilde{C}_2 \oplus \tilde{C}_3)$};
\node[rule, below=0.18cm of r2] (r3) {$f_3$: $\beta_3 \cdot \sum_{i=1}^{4} \alpha_i\tilde{C}_i$};

\node[label, above=0.06cm of r1.north] {Firing Strength};

\node[aggregate, right=0.95cm of r2] (agg) {Weighted\\Sum};
\node[output, right=0.70cm of agg] (out) {Fatigue\\Score};

\draw[dataarrow] (eye.east) -- (norm_eye.west);
\draw[dataarrow] (fnirs.east) -- (norm_fnirs.west);
\draw[dataarrow] (norm_eye.east) -- (eye_enc.west);
\draw[dataarrow] (norm_fnirs.east) -- (fnirs_enc.west);

\draw[arrow] (eye_enc.east) -- ++(0.25,0) |- (c1.west);
\draw[arrow] (eye_enc.east) -- ++(0.25,0) |- (c2.west);
\draw[arrow] (fnirs_enc.east) -- ++(0.25,0) |- (c3.west);
\draw[arrow] (fnirs_enc.east) -- ++(0.25,0) |- (c4.west);
\draw[arrow, dashed, gray!50] (eye_enc.east) -- ++(0.18,0) |- ([yshift=0.06cm]c4.west);

\draw[arrow] (c1.east) -- ++(0.30,0) |- (r1.west);
\draw[arrow] (c2.east) -- ++(0.20,0) |- (r2.west);
\draw[arrow] (c3.east) -- ++(0.20,0) |- (r2.west);
\draw[arrow] (c4.east) -- ++(0.30,0) |- (r3.west);

\draw[arrow, gray!40] (c1.east) -- ++(0.40,0) |- ([yshift=0.05cm]r3.west);
\draw[arrow, gray!40] (c2.east) -- ++(0.30,0) |- (r3.west);
\draw[arrow, gray!40] (c3.east) -- ++(0.30,0) |- ([yshift=-0.05cm]r3.west);

\draw[arrow] (r1.east) -- ++(0.18,0) |- ([yshift=0.12cm]agg.west);
\draw[arrow] (r2.east) -- (agg.west);
\draw[arrow] (r3.east) -- ++(0.18,0) |- ([yshift=-0.12cm]agg.west);

\draw[dataarrow] (agg.east) -- (out.west);

\node[label, above=0.35cm of eye.north] {Input};
\node[label, above=0.35cm of norm_eye.north] {Normalize};
\node[label, above=0.35cm of eye_enc.north] {Encode};
\node[label, above=0.30cm of c1.north] {Concepts};
\node[label, above=0.35cm of agg.north] {Output};

\end{tikzpicture}
}%
\caption{Neuro-symbolic architecture: participant-normalized features $\to$ concept bottleneck (C1--C4) $\to$ differentiable Logic Rules $\to$ fatigue score.}
\label{fig:arch}
\vspace{-0.6em}
\end{figure*}
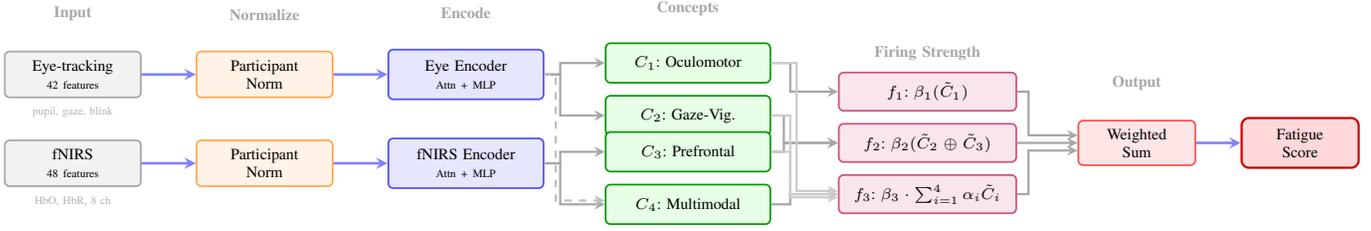

The model predicts binary fatigue $y\in\{0,1\}$ from 10\,s windows (50\% overlap) of eye-tracking and fNIRS. We extract 90 features (42 eye, 48 fNIRS) capturing oculomotor dynamics and prefrontal hemodynamics, then apply participant-aware normalization before an attention-based concept extractor and a differentiable approximate reasoning rule layer (Fig.~\ref{fig:arch}).

\begin{table}[t]
\centering
\scriptsize
\setlength{\tabcolsep}{2pt}
\renewcommand{\arraystretch}{1.05}
\caption{Summary of extracted multimodal features (90 dims total).} 
\label{tab:feature_defs}
\resizebox{0.9\columnwidth}{!}{
\begin{tabularx}{\linewidth}{l X c}
\toprule
\textbf{Modality} & \textbf{Feature Groups \& Metrics} & \textbf{Dim.} \\
\midrule
\textbf{Eye: Pupil} & \textbf{Stat:} $\mu, \sigma$, range, skew, kurtosis. \textbf{Dyn:} $\mu, \sigma, \max$ of $|\dot{p}|$ and $|\ddot{p}|$. \textbf{Spec:} Bandpower (LF, HF, Ratio), SampEn, Coeff.~Var & 16 
\\
\midrule
\textbf{Eye: Oculomotor} & \textbf{Dispersion:} $\sigma_x, \sigma_y$, corr$(x,y)$, spatial $H$. \textbf{Kinematics:} Velocity $v$ ($\mu, \sigma, \max, P_{90}$), accel.~stats. \textbf{Events:} Saccade rate, fixation prop. \textbf{Trend:} Linear fit slope $(x,y)$, angular $\Delta$ stats, SampEn$(v)$. & 18 \\
\midrule
\textbf{Eye: Eyelid} & \textbf{Blink:} Rate, duration/IBI stats ($\mu, \sigma, \min/\max$), Percentage of Eye Closure (PERCLOS) (total \& weighted). & 8 \\
\midrule
\textbf{fNIRS} & \textbf{Global:} $\mu, \sigma$, skew, range of channel-mean \& deriv. \textbf{Spec:} VLF, LF, HF powers \& ratios. \textbf{Regional:} $\mu, \sigma$, SampEn per ROI (8 groups). \textbf{Sym:} L/R diff stats ($\mu, \sigma, \nabla$), corr(L,R), A/P contrast. \textbf{Cmplx:} Global SampEn, Hurst exp., outlier prop. & 48 \\
\bottomrule
\end{tabularx}}
\end{table}


\subsection{Participant-Aware Normalization}

To address the high inter-subject variability characteristic of physiological computing~\cite{fairclough2009physiological}, we normalize features relative to each participant's pre-task alert state. For participant $p$, we compute baseline statistics from samples labeled as alert ($y = 0$):
\begin{equation}
\mu_p = \frac{1}{|S_p^0|} \sum_{i \in S_p^0} \mathbf{x}_i, \quad 
\sigma_p = \text{std}(\{\mathbf{x}_i\}_{i \in S_p^0})
\label{eq:normz}
\end{equation}
where $S_p^0$ denotes the set of alert samples for participant $p$. Normalized features represent relative change from baseline:
$\tilde{\mathbf{x}} = (\mathbf{x} - \mu_p) / (\sigma_p + \epsilon)$,
where $\epsilon$ is a numerical stability constant.

\textbf{Evaluation note}: Pre-task alert segments are included in the held-out subject's evaluation set; they are only used to estimate $\mu_p$ and $\sigma_p$ (normalization), not to tune model weights. While global normalization achieved higher accuracy in ablation studies (80.3\% vs 72.1\%), it assumes cohort-estimated population statistics and is less robust to subject-specific shifts and domain shift. We report participant-aware results as the primary configuration to reflect realistic deployment conditions.

\subsection{Concept Extraction with Learned Attention}

We extract four concepts capturing distinct physiological aspects of fatigue, following the concept bottleneck paradigm~\cite{koh2020concept}:

\begin{enumerate}
\item \textbf{Oculomotor Dynamics} ($C_1$): Encodes gaze kinematics (acceleration, angular velocity, dispersion). It correlates most with gaze acceleration ($r=0.43$) and strongly with fatigue labels ($r=0.86$), consistent with degraded oculomotor control under fatigue.

\item \textbf{Gaze Stability} ($C_2$): Captures compensatory oculomotor patterns (gaze slope, angular dynamics). It is negatively correlated with fatigue ($r=-0.65$), suggesting vigilance-maintenance mechanisms that diminish as fatigue increases.

\item \textbf{Prefrontal Hemodynamics} ($C_3$): Represents prefrontal activity from fNIRS channels and is negatively correlated with fatigue ($r=-0.68$), consistent with reduced prefrontal activation in fatigued states~\cite{causse2017mental}.

\item \textbf{Multimodal} ($C_4$): Models cross-modal eye–fNIRS interactions with a positive fatigue correlation ($r=0.70$).
\end{enumerate}

We interpret concepts via post-hoc correlations between concept activations and input features over the full dataset. Notably, $C_1$ emphasizes gaze kinematics rather than pupil features despite both being available, indicating stronger discriminative value of dynamics in our protocol. This illustrates an advantage of learned concepts over hand-crafted definitions.

Each concept extractor applies learned attention~\cite{niu2021attention} to focus on discriminative features:
\begin{align}
\mathbf{a} &= \sigma(\mathbf{W}_a \tilde{\mathbf{x}}), \quad
\mathbf{h} = \text{GELU}(\text{LN}(\mathbf{W}_h (\tilde{\mathbf{x}} \odot \mathbf{a}))) \\
C_i &= \sigma(\mathbf{v}_i^\top \mathbf{h} + b_i)
\end{align}
where $\sigma$ denotes the sigmoid function, $\odot$ element-wise multiplication, LN is LayerNorm, and concept activations $C_i \in [0,1]$ are interpreted as membership degrees. The hidden dimension is 64 with dropout rate 0.3.

\subsection{Differentiable Approximate Reasoning with Learned Weights}
The neuro-symbolic layer learns concept thresholds ($\tau_i$) and rule weights ($\beta_j$) end-to-end. It also applies an attention-like weighting over concepts ($\boldsymbol{\alpha}$) \emph{within the logic layer}, improving robustness to high inter-subject variability in physiological signals. By learning soft thresholds and rule strengths from data, the model avoids the rigidity of fixed hand-crafted rules~\cite{marra2024nesy}.

We apply soft thresholding with learnable parameters:
\begin{equation}
\tau_i = \sigma(\hat{\tau}_i), \qquad
\tilde{C}_i = \sigma\left((C_i - \tau_i) \cdot T\right)
\label{eq:threshold}
\end{equation}
where $\hat{\tau}_i$ is the learned threshold parameter (initialized to 0.0 so that $\tau_i$ starts at 0.5) and $T = 2.0$ is the temperature controlling decision sharpness. Each concept $C_i \in [0,1]$ is transformed into a soft-thresholded activation $\tilde{C}_i$ centered around its learned threshold. The model handles inverted feature-label relationships through end-to-end learning.

Three rules combine the soft-thresholded concepts $\tilde{C}_i$ (which are already in $[0,1]$) using logic operations, as shown in Table~\ref{tab:rules}. In particular, $\boldsymbol{\alpha}$ denotes \emph{logic-layer concept weights} used only in the global evidence rule $R_3$; we parameterize $\boldsymbol{\alpha}$ by unconstrained logits $\mathbf{w}_\alpha \in \mathbb{R}^4$ and map them to a simplex with softmax:
\begin{equation}
\boldsymbol{\alpha} = \mathrm{softmax}(\mathbf{w}_\alpha),
\end{equation}
so that $\alpha_i \in (0,1)$ and $\sum_{i=1}^{4}\alpha_i = 1$. Larger $\alpha_i$ increases the contribution of $\tilde{C}_i$ to the aggregation in $R_3$.

\begin{table}[ht]
\centering
\caption{Differentiable approximate reasoning rule base with learned weights. Each rule maps concept activations to fatigue evidence.}
\label{tab:rules}
\setlength{\tabcolsep}{3pt}
\renewcommand{\arraystretch}{1.15}
\resizebox{1.01\columnwidth}{!}{
\begin{tabular}{@{}l l l@{}}
\toprule
\textbf{Rule} & \textbf{Antecedent $\rightarrow$ Consequent} & \textbf{Firing Strength} \\
\midrule
$R_1$ &
IF $\tilde{C}_1$ is HIGH THEN Fatigue is LIKELY &
$f_1 = \beta_1 \cdot \tilde{C}_1$ \\

$R_2$ &
IF ($\tilde{C}_2$ is HIGH) OR ($\tilde{C}_3$ is HIGH) THEN Fatigue is LIKELY &
$f_2 = \beta_2 \cdot (\tilde{C}_2 \oplus \tilde{C}_3)$ \\

$R_3$ &
IF (Weighted Concept Sum) is HIGH THEN Fatigue is LIKELY &
$f_3 = \beta_3 \cdot \left(\sum_{i=1}^{4}\alpha_i\tilde{C}_i\right)$ \\
\bottomrule
\end{tabular}}

\begin{flushleft}
\scriptsize

$\oplus$: t-conorm; default probabilistic sum $a \oplus b = a + b - ab$.

$\tilde{C}_i = \sigma((C_i - \tau_i) \cdot T)$: soft-thresholded activation (Eq.~\ref{eq:threshold}).

$\alpha_i$: learned concept weights ($\sum_i \alpha_i=1$); $\beta_j$: learned rule weights.

$\hat{y} = \sigma(\mathbf{w}^\top [f_1,f_2,f_3] + b)$: consequent aggregation.

\end{flushleft}
\end{table}

Post-hoc analysis revealed differing concept polarities: $C_1$ correlates positively with fatigue ($r=0.86$), while $C_2$ and $C_3$ correlate negatively ($r=-0.65$ and $r=-0.68$, respectively). Rule $R_2$ uses the probabilistic-sum t-conorm, $a \oplus b = a + b - ab$, which is smooth and bounded in $[0,1]$. Rule weights $\boldsymbol{\beta}$ are learned during training. The final prediction combines rule outputs through a learned linear layer and sigmoid, adapting to concept polarities without requiring manual specification of semantics.

\subsection{Training Objective}

The total loss combines classification and interpretability regularization:
\begin{equation}
\mathcal{L} = \mathcal{L}_{\text{CE}} + \lambda_1 \mathcal{L}_{\text{div}} + \lambda_2 \mathcal{L}_{\text{sparse}}
\end{equation}
where $\mathcal{L}_{\text{CE}}$ is binary cross-entropy loss. $\mathcal{L}_{\text{div}}$ penalizes correlations between different concepts to encourage disentanglement. $\mathcal{L}_{\text{sparse}}$ promotes interpretable logic by encouraging low-entropy concept weights $\boldsymbol{\alpha}$, concentrating importance on fewer concepts. We set $\lambda_1 = 0.05$ and $\lambda_2 = 0.001$. Training uses AdamW with learning rate $5\times 10^{-4}$, weight decay $10^{-3}$, batch size 32, cosine annealing, gradient clipping at 1.0, and early stopping (patience 20) for up to 150 epochs.

\subsection{Concept Fidelity}

We define concept fidelity $\Phi$ as a post-hoc, label-dependent audit metric. For subject $s$, fidelity is computed as the average absolute point-biserial correlation between concept activations and ground-truth labels:
\begin{equation}
\Phi^{(s)} = \frac{1}{k}\sum_{i=1}^{k}\left|r_{pb}(C_i^{(s)}, y^{(s)})\right|
\label{eq:fidelity}
\end{equation}
Rule fidelity is defined analogously using rule activations $R_j$. High fidelity indicates strong concept-label alignment for a given subject, while low fidelity signals potential subject–model mismatch.

\section{Results}

\subsection{Classification Performance}

Table~\ref{tab:results} proposes LOSO-CV accuracy across all methods. Our neuro-symbolic approach achieves 72.1\% $\pm$ 12.3\%, competitive with both tuned and default baselines while providing interpretable decision pathways. A Wilcoxon signed-rank test comparing NeSy to the best baseline (SVM-RBF tuned) yielded $W=56$, $p=0.332$, with matched-pairs rank-biserial correlation $r=0.27$ (small-to-medium effect); the non-significant result reflects comparable accuracy, with NeSy providing the added benefit of interpretability.

\begin{table}[th]
\centering
\caption{Classification Accuracy (LOSO Cross-Validation)}
\label{tab:results}
\resizebox{0.8\columnwidth}{!}{
\begin{tabular}{@{}lcc@{}}
\toprule
\textbf{Method} & \textbf{Accuracy (\%)}$^\dagger$ & \textbf{95\% CI} \\
\midrule
\textbf{NeSy (Ours)} & \textbf{72.1 $\pm$ 12.3} & [66.4, 77.8] \\
No Logic (Ablation) & 72.0 $\pm$ 12.1 & [66.4, 77.6] \\
\midrule
\multicolumn{3}{@{}l@{}}{\textit{Tuned Baselines (nested CV):}} \\
SVM-RBF & 69.4 $\pm$ 13.5 & [63.2, 75.7] \\
Random Forest & 66.6 $\pm$ 12.3 & [60.9, 72.3] \\
Extra Trees & 68.3 $\pm$ 14.3 & [61.7, 74.9] \\
Logistic Regression & 67.8 $\pm$ 15.1 & [60.8, 74.7] \\
\midrule
\multicolumn{3}{@{}l@{}}{\textit{Default Baselines:}} \\
SVM-RBF & 67.6 $\pm$ 16.4 & [60.0, 75.2] \\
Random Forest & 64.7 $\pm$ 13.6 & [58.4, 71.0] \\
\bottomrule
\multicolumn{3}{@{}l@{}}{\scriptsize $^\dagger$Mean $\pm$ SD across 18 LOSO folds (seed-averaged per fold); 95\% CI across folds}
\end{tabular}}
\end{table}

\subsection{Ablation Studies}

We conducted ablation studies to isolate the contribution of each architectural component.

\subsubsection{Normalization Strategies}

Table~\ref{tab:ablation_norm} compares normalization approaches under realistic deployment conditions. Global normalization (using training-cohort statistics) achieves 80.3\%, but it assumes that cohort-level feature distributions transfer to unseen users. Our participant-aware calibration approach achieves 72.1\%. Removing calibration entirely drops accuracy to 66.9\%, demonstrating a +5.2 pp absolute improvement.

\begin{table}[ht]
\centering
\caption{Ablation: Normalization Strategies}
\label{tab:ablation_norm}
\resizebox{0.9\columnwidth}{!}{
\begin{tabular}{@{}lcc@{}}
\toprule
\textbf{Strategy} & \textbf{Accuracy (\%)} & \textbf{Std (\%)} \\
\midrule
Global$^*$ & 80.3 & 10.1 \\
Participant-Aware (Calibration)$^\dagger$ & 72.1 & 12.3 \\
No-Calibration (Train-Only) & 66.9 & 8.3 \\
\bottomrule
\multicolumn{3}{@{}l@{}}{\scriptsize $^*$Population statistics estimated from the training cohort; may not transfer to unseen users}\\
\multicolumn{3}{@{}l@{}}{\scriptsize $^\dagger$Primary configuration; calibration, no test labels used}
\end{tabular}}
\end{table}

\subsubsection{Individual Concept Contributions}

Table~\ref{tab:ablation_concepts} shows the accuracy drop when each concept is ablated (zeroed out) under the calibration configuration. The prefrontal-hemodynamic concept ($C_3$) contributes most significantly, with a 1.2 pp accuracy drop when removed. This finding aligns with the rule discrimination analysis showing $R_2$ (which incorporates $C_3$) as the dominant discriminator.

\begin{table}[ht]
\centering
\caption{Ablation: Individual Concept Contributions}
\label{tab:ablation_concepts}
\resizebox{0.9\columnwidth}{!}{
\begin{tabular}{@{}lcc@{}}
\toprule
\textbf{Configuration} & \textbf{Accuracy (\%)} & \textbf{$\Delta$ (pp)} \\
\midrule
Full Model (Calibration) & 72.1 & -- \\
Without $C_1$ (Oculomotor Dynamics) & 73.4 & +1.3 \\
Without $C_2$ (Gaze Stability) & 72.4 & +0.3 \\
Without $C_3$ (Prefrontal Hemodynamics) & 70.9 & -1.2 \\
Without $C_4$ (Multimodal) & 71.2 & -0.9 \\
\bottomrule
\end{tabular}}
\end{table}

Post-hoc correlation analysis revealed that $C_1$ learned to weight gaze acceleration and angular dynamics rather than pupil-specific features, despite both being available in the input. This suggests that oculomotor control degradation provided a stronger discriminative signal than pupil dynamics in our protocol. The modest accuracy improvement when removing
$C_1$ ($+1.3$ pp) and $C_2$ ($+0.3$ pp) indicate partial redundancy among eye-tracking concepts, while removing $C_3$ (Prefrontal-hemodynamic) causes the largest drop ($-1.2$ pp), consistent with $R_2$ being the dominant discriminator.

\subsubsection{Logic Operators}

Table~\ref{tab:ablation_tnorm} compares different logic operator families under the calibration configuration. The ``Product'' configuration uses product t-norm ($a \cdot b$) for conjunction and probabilistic sum ($a + b - ab$) for disjunction in $R_2$. ``{\L}ukasiewicz'' uses bounded operators: $\max(0, a+b-1)$ for conjunction and $\min(1, a+b)$ for disjunction. ``G\"{o}del'' uses $\min/\max$ operations. {\L}ukasiewicz operators achieve the highest accuracy (72.9\%).

\begin{table}[ht]
\centering
\caption{Ablation: Logic Operators}
\label{tab:ablation_tnorm}
\resizebox{0.8\columnwidth}{!}{
\begin{tabular}{@{}lccc@{}}
\toprule
\textbf{Operator} & \textbf{Accuracy (\%)} & \textbf{Std (\%)} & \textbf{$\Delta$} \\
\midrule
Product & 72.0 & 12.0 & -- \\
{\L}ukasiewicz & \textbf{72.9} & 12.2 & +0.9 \\
G\"{o}del (min/max) & 72.2 & 12.3 & +0.2 \\
\bottomrule
\end{tabular}}
\end{table}

\subsubsection{Learned versus Fixed Thresholds}

Table~\ref{tab:ablation_polarity} compares learned thresholds $\tau_i$ against fixed values ($\tau_i = 0.5$ for all concepts) under the calibration configuration. Learned thresholds provide a modest improvement (+1.5 pp), allowing the model to adapt decision boundaries to each concept's distribution. The similar performance suggests the model is relatively robust to threshold initialization.

\begin{table}[ht]
\centering
\caption{Ablation: Learned vs. Fixed Thresholds}
\label{tab:ablation_polarity}
\begin{tabular}{@{}lcc@{}}
\toprule
\textbf{Configuration} & \textbf{Accuracy (\%)} & \textbf{Std (\%)} \\
\midrule
Learned Thresholds & \textbf{72.1} & 12.3 \\
Fixed Thresholds ($\tau=0.5$) & 70.6 & 12.2 \\
\bottomrule
\end{tabular}
\end{table}

\subsection{Concept Fidelity and Rule Discrimination}

Table~\ref{tab:fidelity} presents per-subject fidelity and rule discrimination metrics. Mean concept fidelity was $\Phi = 0.41 \pm 0.13$, with strong correlation with classification accuracy ($r = 0.843$, $p < 10^{-4}$).

\begin{table}[ht]
\centering
\caption{Per-subject fidelity and rule discrimination}
\label{tab:fidelity}
\resizebox{0.7\linewidth}{!}{%
\begin{tabular}{@{}lcccccc@{}}
\toprule
\textbf{ID} & \textbf{Acc} & \textbf{C-Fid} & \textbf{R-Fid} & \textbf{$R_1$} & \textbf{$R_2$} & \textbf{$R_3$} \\
\midrule
P007 & 96.7 & 0.75 & 0.83 & -0.03 & +0.09 & -0.01 \\
P006 & 86.4 & 0.55 & 0.56 & -0.05 & +0.10 & -0.00 \\
P010 & 85.0 & 0.46 & 0.48 & -0.05 & +0.10 & -0.00 \\
P018 & 85.0 & 0.44 & 0.61 & -0.04 & +0.06 & -0.01 \\
P011 & 83.5 & 0.45 & 0.44 & -0.04 & +0.08 & -0.00 \\
P016 & 81.4 & 0.48 & 0.47 & +0.01 & +0.05 & +0.00 \\
\midrule
P009 & 54.4 & 0.25 & 0.27 & +0.00 & +0.02 & +0.00 \\
P005 & 55.3 & 0.14 & 0.10 & -0.01 & +0.01 & -0.00 \\
P002 & 58.0 & 0.39 & 0.25 & +0.01 & +0.02 & +0.01 \\
P001 & 58.3 & 0.16 & 0.28 & -0.03 & +0.01 & -0.00 \\
\midrule
\textbf{Mean} & 72.1 & 0.41 & 0.40 & -0.02 & +0.05 & -0.00 \\
\bottomrule
\end{tabular}
}
\begin{flushleft}
\scriptsize C-Fid: concept fidelity; R-Fid: rule fidelity; $R_1$--$R_3$: rule discrimination (Fatigued$-$Alert). Ten representative subjects shown.
\end{flushleft}
\end{table}

Rule discrimination analysis reveals key patterns:
\begin{itemize}
\item \textbf{$R_2$ (Vigilance \& Prefrontal)}:
Mean discrimination $+0.052$, Cohen's $d = 1.15$ (large effect). 
This rule combines gaze stability and prefrontal haemodynamics via probabilistic sum
is the primary discriminator.
\item \textbf{$R_1$ (Oculomotor Dynamics)}: Mean discrimination $-0.021$, Cohen's $d = -0.24$. The small negative discrimination at the rule level reflects the interaction between the learned concept ($C_1$, which correlates positively with fatigue at $r=0.86$) and the learned rule weight $\beta_1$, which the model adjusted to balance contributions across rules.
\item \textbf{$R_3$ (Combined)}: Near-zero discrimination ($-0.001$), serving as a regularizing baseline rather than active discriminator.
\end{itemize}

Concept fidelity is computed \emph{post-hoc} from ground-truth labels (Eq.~\ref{eq:fidelity}), and thus serves as an \emph{offline} alignment diagnostic (useful in validation/pilot settings with occasional labels) to flag subjects needing re-calibration or sensor/assessment checks. Developing a label-free reliability proxy that tracks fidelity remains future work.

\subsection{Per-Subject Analysis}

The best-performing subject (P007, 96.7\% accuracy) achieved near-perfect classification with high concept fidelity (0.75) and strong $R_2$ discrimination (+0.09). Several subjects (P009, P005, P001, P002) performed below 60\% with low concept fidelity and weak rule discrimination, suggesting poor concept--label alignment for these individuals.

\begin{figure}[t]
    \centering
    \includegraphics[width=0.75\linewidth]{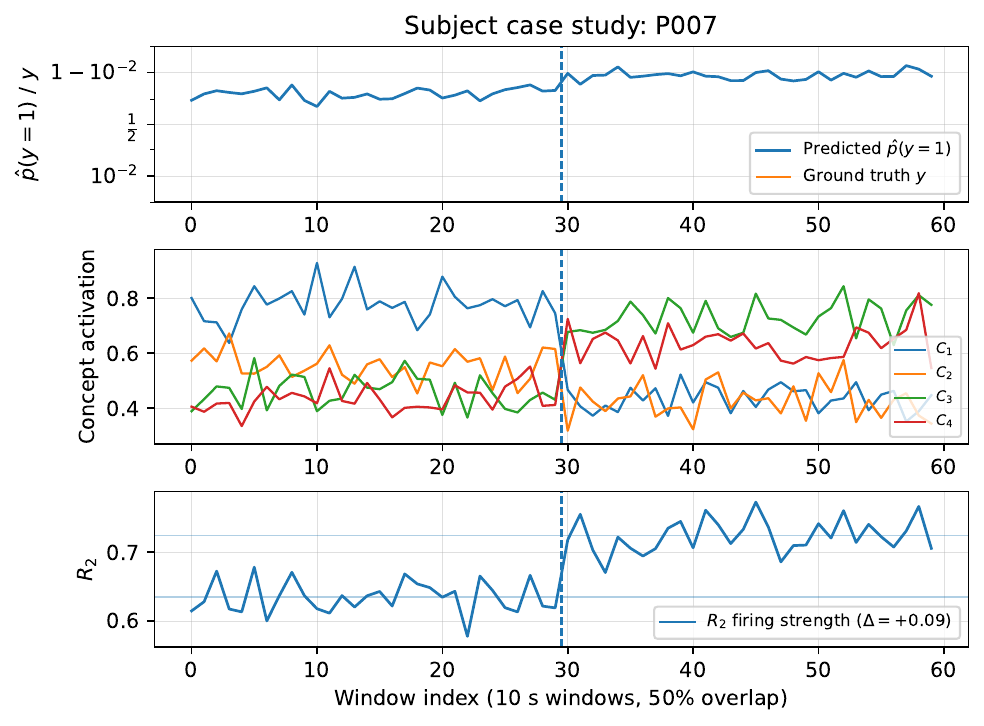}
    \caption{Case study (P007): fatigue prediction and labels (top), concept activations $C_1$--$C_4$ (middle), and rule $R_2$ firing (bottom); dashed line marks the Alert$\rightarrow$Fatigued transition.}
    \label{fig:case_p007}
\end{figure}

We also observe that a subset of participants exhibit weak or even reversed rule discrimination, suggesting that the effective direction of some fatigue-related patterns can vary across individuals; our rule-level diagnostics make such subject-specific deviations explicit.

\section{Discussion}
\subsection{Interpretability versus Performance Trade-off}

The full model (72.1\%) matches the no-logic ablation (72.0\%), indicating the approximate reasoning layer primarily improves interpretability rather than accuracy. This is intentional: we trade marginal gains for transparent decision paths, enabling inspection of which physiological cues drive predictions and whether errors arise from poor signal quality, genuine ambiguity, or subject-specific fatigue patterns.

\subsection{Ablation Study Insights}

Ablations yield three key findings. Removing the prefrontal hemodynamics concept ($C_3$) and multimodal fusion ($C_4$) drops accuracy by 1.2 pp and 0.9 pp, respectively, while removing gaze-vigilance ($C_2$) and oculomotor-fatigue ($C_1$) slightly increases accuracy (+0.3 pp, +1.3 pp), suggesting redundancy or noise sensitivity in gaze-derived metrics. {\L}ukasiewicz pooling performs best (72.9\%), exceeding product operators by 0.9 pp. Finally, learned thresholds improve accuracy by 1.5 pp over fixed $\tau=0.5$, highlighting the value of adaptive decision boundaries.

\subsection{Limitations}

Sample size ($N = 18$) limits generalizability. The labeling strategy conflates fatigue with time-on-task effects. The logic layer provides interpretability without significant accuracy improvement over non-interpretable alternatives. Future work should validate on larger cohorts, explore temporal dynamics, and investigate online adaptation for low-fidelity subjects.

\section{Conclusion}

We introduced a neuro-symbolic fatigue detection framework that learns interpretable oculomotor and hemodynamic concepts from eye-tracking and fNIRS and fuses them via differentiable approximate reasoning. Key findings are:

\begin{enumerate}
\item Competitive performance: 72.1\% $\pm$ 12.3\% LOSO-CV accuracy with interpretable decision pathways.
\item Ablations show calibration is most impactful (+5.2 pp), while fNIRS contributes (+1.2 pp) and {\L}ukasiewicz operators outperform product (+0.9 pp).
\item Concept fidelity correlates strongly with accuracy ($r=0.843$, $p<10^{-4}$), enabling post-hoc reliability auditing.
\item Rule analysis identifies $R_2$ as the main discriminator (Cohen's $d=1.15$) and reveals subject-specific patterns missed by black-box models.
\end{enumerate}

Finally, the framework delivers transparent, expert-auditable fatigue monitoring alongside clearer indicators of prediction trustworthiness.

\section{Acknowledgement}

This research and Javier Fumanal-Idocin were supported by EU Horizon Europe under the Marie Skłodowska-Curie COFUND grant No 101081327 YUFE4Postdocs. This research is supported by UKRI BBSRC project EyeWarn (code: APP37953).

\bibliographystyle{IEEEtran}
\bibliography{paper_wcci2026_6pages}

@article{caldwell2019fatigue,
  title={Fatigue and its management in the workplace},
  author={Caldwell, John A and Caldwell, J Lynn and Thompson, Lauren A and Lieberman, Harris R},
  journal={Neuroscience \& Biobehavioral Reviews},
  volume={96},
  pages={272--289},
  year={2019},
  publisher={Elsevier}
}

@article{ajayi2025multimodal,
  title={A Multimodal Systematic Review of Drivers’ Fatigue Detection Methodologies, Datasets, and Models},
  author={Ajayi, OO and Kurien, AM and Djouani, K and Dieng, L},
  journal={IEEE Access},
  year={2025},
  publisher={IEEE}
}

@article{kashevnik2024gaze,
  title={Intelligent Human Operator Mental Fatigue Assessment Method Based on Gaze Movement Monitoring},
  author={Kashevnik, Alexey and Kovalenko, Svetlana and Mamonov, Anton and Hamoud, Batol and Bulygin, Aleksandr and Kuznetsov, Vladislav and Shoshina, Irina and Brak, Ivan and Kiselev, Gleb},
  journal={Sensors},
  volume={24},
  number={21},
  pages={6805},
  year={2024},
  publisher={MDPI}
}

@article{ayaz2012optical,
  title={Optical brain monitoring for operator training and mental workload assessment},
  author={Ayaz, Hasan and Shewokis, Patricia A and Bunce, Scott and Izzetoglu, Kurtulus and Willems, Ben and Onaral, Banu},
  journal={Neuroimage},
  volume={59},
  number={1},
  pages={36--47},
  year={2012},
  publisher={Elsevier}
}

@article{causse2017mental,
  title={Mental workload and neural efficiency quantified in the prefrontal cortex using fNIRS},
  author={Causse, Micka{\"e}l and Chua, Zarrin and Peysakhovich, Vsevolod and Del Campo, Natalia and Matton, Nadine},
  journal={Scientific reports},
  volume={7},
  number={1},
  pages={5222},
  year={2017},
  publisher={Nature Publishing Group UK London}
}

@article{hossain2023bci,
  title={Status of deep learning for EEG-based brain--computer interface applications},
  author={Hossain, Khondoker Murad and Islam, Md Ariful and Hossain, Shahera and Nijholt, Anton and Ahad, Md Atiqur Rahman},
  journal={Frontiers in computational neuroscience},
  volume={16},
  pages={1006763},
  year={2023},
  publisher={Frontiers Media SA}
}

@article{marra2024nesy,
  title={From statistical relational to neurosymbolic artificial intelligence: A survey},
  author={Marra, Giuseppe and Duman{\v{c}}i{\'c}, Sebastijan and Manhaeve, Robin and De Raedt, Luc},
  journal={Artificial Intelligence},
  volume={328},
  pages={104062},
  year={2024},
  publisher={Elsevier}
}

@misc{colelough2025nesy,
  title={Neuro-symbolic AI in 2024: A systematic review},
  author={Colelough, Brandon C and Regli, William},
  journal={arXiv preprint arXiv:2501.05435},
  year={2025}
}

@incollection{badreddine2021logic,
  title={Logic tensor networks: Theory and applications},
  author={Serafini, Luciano and Badreddine, Samy and Donadello, Ivan and Spranger, Michael and Bianchi, Federico and others},
  booktitle={Neuro-Symbolic Artificial Intelligence: The State of the Art},
  pages={370--394},
  year={2021},
  publisher={IOS Press}
}

@inproceedings{koh2020concept,
  title={Concept bottleneck models},
  author={Koh, Pang Wei and Nguyen, Thao and Tang, Yew Siang and Mussmann, Stephen and Pierson, Emma and Kim, Been and Liang, Percy},
  booktitle={International conference on machine learning},
  pages={5338--5348},
  year={2020},
  organization={PMLR}
}

@article{yuksekgonul2022posthoc,
  title={Post-hoc concept bottleneck models},
  author={Yuksekgonul, Mert and Wang, Maggie and Zou, James},
  journal={arXiv preprint arXiv:2205.15480},
  year={2022}
}

@article{kaya2024eeg,
  title={EEG-based emotion recognition in neuromarketing using fuzzy linguistic summarization},
  author={Kaya, {\"U}mran and Akay, Diyar and Ayan, Sevgi {\c{S}}eng{\"u}l},
  journal={IEEE Transactions on Fuzzy Systems},
  volume={32},
  number={8},
  pages={4248--4259},
  year={2024},
  publisher={IEEE}
}

@article{bhuyan2024nesy,
  title={Neuro-symbolic artificial intelligence: a survey},
  author={Bhuyan, Bikram Pratim and Ramdane-Cherif, Amar and Tomar, Ravi and Singh, TP},
  journal={Neural Computing and Applications},
  volume={36},
  number={21},
  pages={12809--12844},
  year={2024},
  publisher={Springer}
}

@article{rudin2019stop,
  title={Stop explaining black box machine learning models for high stakes decisions and use interpretable models instead},
  author={Rudin, Cynthia},
  journal={Nature machine intelligence},
  volume={1},
  number={5},
  pages={206--215},
  year={2019},
  publisher={Nature Publishing Group UK London}
}

@inproceedings{guo2017calibration,
  title={On calibration of modern neural networks},
  author={Guo, Chuan and Pleiss, Geoff and Sun, Yu and Weinberger, Kilian Q},
  booktitle={International conference on machine learning},
  pages={1321--1330},
  year={2017},
  organization={PMLR}
}

@inproceedings{gal2016dropout,
  title={Dropout as a bayesian approximation: Representing model uncertainty in deep learning},
  author={Gal, Yarin and Ghahramani, Zoubin},
  booktitle={international conference on machine learning},
  pages={1050--1059},
  year={2016},
  organization={PMLR}
}

@article{lakshminarayanan2017simple,
  title={Simple and scalable predictive uncertainty estimation using deep ensembles},
  author={Lakshminarayanan, Balaji and Pritzel, Alexander and Blundell, Charles},
  journal={Advances in neural information processing systems},
  volume={30},
  year={2017}
}

@article{fairclough2009physiological,
  title={Fundamentals of physiological computing},
  author={Fairclough, Stephen H},
  journal={Interacting with computers},
  volume={21},
  number={1-2},
  pages={133--145},
  year={2009},
  publisher={OUP}
}

@article{niu2021attention,
  title={A review on the attention mechanism of deep learning},
  author={Niu, Zhaoyang and Zhong, Guoqiang and Yu, Hui},
  journal={Neurocomputing},
  volume={452},
  pages={48--62},
  year={2021},
  publisher={Elsevier}
}

@article{lei2023end,
  title={An end-to-end review of gaze estimation and its interactive applications on handheld mobile devices},
  author={Lei, Yaxiong and He, Shijing and Khamis, Mohamed and Ye, Juan},
  journal={ACM Computing Surveys},
  volume={56},
  number={2},
  pages={1--38},
  year={2023},
  publisher={ACM New York, NY}
}

@misc{lei5732154comprehensive,
  author = {Lei, Yaxiong and Azizinezhad, Parastoo and Jamalifard, Mohammadreza and Manohar, Sanjay and Wlodarski, Michal and Foulsham, Tom and Andreu-Perez, Javier},
  title  = {Ocular Metrics for Fatigue Assessment: A Survey from Physiology to Machine Learning},
  howpublished = {SSRN},
  note   = {Article no. 5732154},
  year   = {2024}
}

@article{albadawi2022review,
  title={A review of recent developments in driver drowsiness detection systems},
  author={Albadawi, Yaman and Takruri, Maen and Awad, Mohammed},
  journal={Sensors},
  volume={22},
  number={5},
  pages={2069},
  year={2022},
  publisher={MDPI}
}

@article{lei2025quantifying,
  title={Quantifying the impact of motion on 2d gaze estimation in real-world mobile interactions},
  author={Lei, Yaxiong and Wang, Yuheng and Buchanan, Fergus and Zhao, Mingyue and Sugano, Yusuke and He, Shijing and Khamis, Mohamed and Ye, Juan},
  journal={arXiv preprint arXiv:2502.10570},
  year={2025}
}

@article{kakhi2025fatigue,
  title={A Transfer Learning-based Approach for Fatigue Detection through the Fusion of Physiological Signals},
  author={Kakhi, Kourosh and Asgharnezhad, Hamzeh and Khosravi, Abbas and Alizadehsani, Roohallah and Acharya, U Rajendra},
  journal={IEEE Sensors Journal},
  year={2025},
  publisher={IEEE}
}

@article{lohani2019review,
  title={A review of psychophysiological measures to assess cognitive states in real-world driving},
  author={Lohani, Monika and Payne, Brennan R and Strayer, David L},
  journal={Frontiers in human neuroscience},
  volume={13},
  pages={57},
  year={2019},
  publisher={Frontiers Media SA}
}

@article{fumanal2024ex,
  title={Ex-Fuzzy: A library for symbolic explainable AI through fuzzy logic programming},
  author={Fumanal-Idocin, Javier and Andreu-Perez, Javier},
  journal={Neurocomputing},
  volume={599},
  pages={128048},
  year={2024},
  publisher={Elsevier}
}

\end{document}